\documentclass{article}

\usepackage[utf8]{inputenc}

\usepackage{xparse,xifthen}
\usepackage{mathtools} 
\usepackage{amsthm,amsxtra}
\usepackage{tikz}
\usetikzlibrary{cd, intersections, decorations.markings, arrows.meta}
\usepackage{thmtools}

\usepackage{enumitem}
\usepackage{braket}

\usepackage{amsfonts}
\usepackage{mathabx}

\usepackage{url}
\usepackage{hyperref}

\DeclareMathAlphabet{\mathsc}{OT1}{cmr}{m}{sc}

\allowdisplaybreaks

\makeatletter
\newcommand{\pushright}[1]{\ifmeasuring@#1\else\omit\hfill$\displaystyle#1$\fi\ignorespaces}
\newcommand{\pushleft}[1]{\ifmeasuring@#1\else\omit$\displaystyle#1$\hfill\fi\ignorespaces}
\makeatother

\newcommand{\RR}{\mathbb R}
\newcommand{\CC}{\mathbb C}
\newcommand{\NN}{\mathbb N}
\newcommand{\ZZ}{\mathbb Z}

\newcommand{\abs}[1]{\lvert #1 \rvert}

\newcommand{\D}{\mathrm{d}}

\newcommand{\e}{\mathrm{e}}

\DeclareDocumentCommand{\EE}{ O{}mO{} }{
	\ifthenelse{\isempty{#3}}{
		\mathbb E_{#1}\!\left[#2\right]}{
		\mathbb E_{#1}\!\left[#2\,\middle|\,#3\right]}
}

\declaretheorem[
	name=Definition, style=definition%
]{defn}

\declaretheorem[
	name=Remark, style=remark, sibling=defn
]{rk}

\author{
Rodrigo Vargas Le-Bert\thanks{This work was partially funded by Associazione LumbeLumbe. The author expresses his deepest gratitude to Horst Thaler for making it possible.}
}

\date{\normalsize\today}

\title{Renormalization flow fixed points for higher-dimensional abelian gauge fields}

\begin{document}

\begin{titlepage}

\maketitle
\thispagestyle{empty}

\begin{abstract}
A connection modulo gauge symmetry on the trivial principal bundle $M\times G$ is a morphism from the loop group of $M$ into $G$.
Thus, considering only loops around the 2-cells of a distinguished family of progressively refined cellular structures on $M$,
the observable algebra $A$ of an abelian gauge field can be presented as an inductive limit of quotients of polynomial algebras.
In that context, it turns out that the state $\mu_\lambda:A\rightarrow\CC$ of the Yang-Mills field on the sphere
can be written $\mu_\lambda = \mu_0\e^{\lambda L}$
with $\lambda$ an interaction strength parameter, $L:A\rightarrow A$ an explicit second-order partial differential operator and $\mu_0$ the state of an almost surely flat connection.
Extrapolating, we provide analogous states for the case of abelian gauge fields on $\RR^d$.
\end{abstract}

\setcounter{tocdepth}{2}
\tableofcontents
\end{titlepage}

\section{Introduction} \label{intro}

Consider the  observable algebra of an abelian gauge theory on a lattice.
This lattice can be taken to be the set of 1-cells $C^1M$ of a cellular structure $C$ on certain riemannian manifold $M$, which we take to be 2-connected.
Assume, for simplicity, that the gauge group is $\RR$.
Then, writing $x_p$ for the holonomy around a plaquette $p\in C^2M$ and $x_C=\set{x_p|p\in C^2M}$, the observable algebra is
\[ 
A_C = \CC[x_C]/I_C
\] 
where $I_C$ is the ideal generated by the Bianchi identities
\[ 
\sum \langle\partial c,p\rangle x_p = 0,\quad c\in C^3M.
\] 
Here, $\partial c$ is the homological boundary of $c$ and $\langle\partial c,p\rangle\in\set{-1,0,1}$ is defined by $\partial c = \sum\langle\partial c,p\rangle p$.
Given a refinement of the cellular structure $C$, there is a morphism of the corresponding algebras, and the observable algebra of the continuum theory is the inductive limit $A = \injlim A_C$.
Our goal is to construct physically plausible \emph{states} on this algebra (that is, to specify the expected values of its observables).

There is a natural reference state on $A_C$: the quotient $\mu_0:\CC[x_C]/I_C\rightarrow \CC$ sending each $x_p\mapsto 0$. Physically, this is the trivial case in which the connection is almost surely flat.
Now, let us plainly state the main novelty in our approach.
It can be seen that the well-known Yang-Mills measure on the sphere (which is not 2-connected but can still be treated similarily, see~\autoref{Yang-Mills}) reads, exactly at each effective scale $C$,
\[
\mu_\lambda = \mu_0\e^{\lambda L},\quad L:A_C\rightarrow A_C
\]
where $L$ is a second-order differential operator and $\lambda>0$.
Explicitely,
\begin{equation} \label{L on the sphere}
L = \sum a_p\partial_p^2 - \sum a_pa_q\partial_p\partial_q
\end{equation}
where $\partial_p = \partial/\partial x_p$ and $a_p$ is proportional to the area of the plaquette $p$, normalized by $\sum a_p = 1$.
Note how the first term generates a heat kernel measure for the holonomy around each plaquette, whereas the second term accounts for the interaction between holonomies enforced by the Bianchi identity.
We use this as an ansatz for the case $M = \RR^d$, coming up with solutions with a similar structure: one term generating heat kernel measures for the holonomies, plus a second term ensuring the Bianchi identities.
We emphasize that the resulting operators at different scales are compatible; thus, they provide the exact renormalization trajectories of certain abelian gauge fields.
Uniqueness, on the other hand, does not hold.

Perhaps the most important unaddressed question, at this stage, is that of reflection positivity of the resulting state.
We hope that our results will encourage further development of this line of attack on constructive quantum field theory.

\section{Invariant differential operators}

Given a cellular structure $C$ on a 2-connected riemannian manifold $M$,
consider the algebra $A_C = \CC[x_C]/I_C$ defined in \autoref{intro}.
We are interested in certain differential operators of the form
\[
L = \sum a_p\partial_p^2 - \sum b_{pq}\partial_p\partial_q
\]
where $\partial_p = \partial/\partial x_p$. Specifically, we focus on those which are well-defined on $A_C$.
We assume, without loss of generality, that $b_{pq}=b_{qp}$.

Given $c\in C^3M$, let $f_c = \sum\langle\partial c,p\rangle x_p\in \CC[x_C]$.
Since elements of $I_C$ can be written as $\sum f_cg_c$ with the $g_c$'s in $\CC[x_C]$,
in order for $L$ to be well-defined it suffices that $L(f_cg)\in I_C$ for all $g\in\CC[x_C]$. Now, we have
\[
L(fg) = L(f)g+fL(g) + 2\Bigl( \sum a_p\partial_pf\partial_pg - \sum b_{pq}\partial_pf\partial_qg \Bigr),
\]
and it is only the third term that can cause trouble. Putting $f=f_c$, that term reads
\[
2\sum_p \partial_pg \Bigl( \langle\partial c,p\rangle a_p - \sum_q \langle\partial c,q\rangle b_{pq} \Bigr).
\]
Thus, all is well if we impose that
\begin{equation} \label{L passes to the quotient}
\langle\partial c,p\rangle a_p - \sum_q \langle\partial c,q\rangle b_{pq} = 0,\quad p\in C^2M,\ c\in C^3M.
\end{equation}

\begin{rk}
If $M$ is not 2-connected, one must test \autoref{L passes to the quotient} on exact 2-chains, as opposed to just boundaries $\partial c$ with $c\in C^3M$. In particular, for $M=\mathbb S^2$, there is only one such 2-chain and the condition boils down to
$a_p = \sum_q b_{pq}$, which is satisfied by the coefficients $b_{pq} = a_pa_q$ in \autoref{L on the sphere} if one assumes that $\sum a_p = 1$.
\end{rk}

Now, take a cellular decomposition $C'$ of $M$ which is finer than $C$ and consider a differential operator
\[
L' = \sum a_{p'}\partial_{p'} - \sum b_{p'q'}\partial_{p'}\partial_{q'},
\]
where $p',q'$ stand for plaquettes in $C'M$.
We are interested in characterizing the case in which $L$ and $L'$ are compatible, in the sense of fitting into a commutative square
\[
\begin{tikzcd}
A_C \ar[r,hook]\ar[d,"L" left] &A_{C'} \ar[d,"L'"] \\
A_C \ar[r,hook] &A_{C'}.
\end{tikzcd}
\]
In order to spell this out note that, assuming without loss of generality that the orientations of plaquettes in $C'$ have been chosen compatible with those in $C$, the inclusion $A_C\hookrightarrow A_{C'}$ is induced by
\[
x_p \mapsto \sum_{p'\subseteq p} x_{p'},\quad p\in C^2M.
\]
Thus, the compatibility condition boils down to
\begin{equation} \label{compatibility}
a_p = \sum_{p'\subseteq p} a_{p'},\quad b_{pq} = \sum_{p'\subseteq p}\sum_{q'\subseteq q} b_{p'q'}.
\end{equation}

\begin{rk}
If $M=\mathbb S^2$,
compatibility between $L$'s at different scales holds for $a_p$ equal to the area of $p$ and $b_{pq} = a_pa_q$ (assuming $\mathbb S^2$ has total area 1).
\end{rk}

\section{Renormalization flow fixed points}


We consider cellular decompositions $C_n$ of $M=\RR^d$ such that $C_n^dM$ consists of hipercubes with sidelength $2^{-n+1}$ and vertices on $C_n^0M = (2^{-n+1}\ZZ)^d$, for some fixed $n\in\NN$.
We need to fix an orientation for the plaquettes $p\in C_n^2M$,
and we do so by simply following the canonical order of the canonical basis vectors.
This choice is arguably not the most natural one from a geometric perspective in three dimensions, but its algebraic simplicity is helpful in higher dimensions.

We will parameterize $C_n$ using points of $\RR^d$,
by letting $[u]_n\in C_n$ be the unique cell containing $u\in\RR^d$.
It is easy to see that $[\cdot]_n$ implements a bijection between $(2^{-n}\ZZ)^d$ and $C_n$.
Now, fix $n=0$. Vertices of $C_{0}$ correspond to vectors $u\in\ZZ^d$ with all coordinates even; edges to vectors with exactly one odd coordinate; faces to vectors with exactly two odd cordinates, and so on.
Next, fix $d=3$ and consider the cube $[1,1,1]$ where, for brevity's sake, we are writing $[u_1,u_2,u_3]$ instead of $[(u_1,u_2,u_3)]_0$. Its homological boundary is
\[
\partial[1,1,1] = -[0,1,1] + [2,1,1] + [1,0,1] - [1,2,1] - [1,1,0] + [1,1,2].
\]
Analogous formulas hold for its translations as well as its orientation-preserving embeddings in $\RR^d$, for $d>3$.

\subsection{$M = \RR^3$} \label{R3}

We set out to specify a compatible family of invariant differential operators
\[
L_n = \sum a_p\partial_p^2 - \sum b_{pq}\partial_p\partial_q.
\]
Here,
the sums are over plaquettes, $a_p$ is proportional to the area of $p$ and $\partial_p$ is the derivative with respect to the variable $x_p$ which, we recall, stands for the holonomy around $p$.

Start with $n=0$.
We impose translation and rotation invariance, thus being allowed to specify only the coefficients $a_0=a_p$ and $b_0(q) = b_{pq}$ with $p=[1,1,0]$ and $q$ contained in the principal octant $\Set{(u_1,u_2,u_3) | u_i\geq 0}$. By furthermore imposing reflection invariance we can just focus on plaquettes which are parallel to (i.e.\ obtainable by translation from) one of
\(
[1,1,0],\ [0,1,1].
\)
Indeed, interchanging two consecutive basis elements reverses the orientation of the plane containing both, leaving all the rest untouched (orientation is a matter of ordering);
thus, under reflection invariance,
\[
b_0([1+2i, 2j, 1+2k]) = -b_0([2j, 1+2i, 1+2k]),\quad i,j,k\in\ZZ
\]
with sign change because the reversed plane contains $p$.
We will write
\begin{align*}
\alpha_0(i,j,k) &= b_0([1+2i,1+2j,2k]) \\
\beta_0(i,j,k) &= b_0([2i,1+2j,1+2k])
\end{align*}
where the subindex 0 is there to remind us that we are working at $n=0$ (and will later consider other values).
Note that, again by reflection invariance, for all $i,j,k\in\ZZ$ one has:
\begin{itemize}
	\item $\beta_0(-i,j,k) = -\beta_0(1+i,j,k)$
 	\item $\beta_0(i,-j,k) = \beta_0(i,j,k)$
	\item $\beta_0(i,j,-k) = -\beta_0(i,j,k)$
\end{itemize}
Assuming that all unspecified coefficients vanish, our choices are as follows.
\begin{enumerate}
\item For $k=0$:
	\begin{enumerate}
		\item $\alpha_0(0,0,0) = 2$
		\item $\beta_0(0,0,0) = -\beta_0(1,0,0) = 2$
		\item $\beta_0(0,1,0) = -\beta_0(1,1,0) = 1$
	\end{enumerate}
\item For $k\geq 1$:
	\begin{enumerate}
		\item $\alpha_0(0,0,k) = \alpha_0(k,k,k) = -2$
		\item $\beta_0(k+1,k,k) = \beta_0(k+1,k+1,k) = -1$
	\end{enumerate}
\end{enumerate}
We proceed to check gauge invariance and multiscale consistency.

Consider the invariance requirement stemming from the Bianchi identity associated to the cube $[1+2i,1+2j,1+2k]$, whose boundary is
\begin{align*}
&-[2i,1+2j,1+2k] + [2i+2,1+2j,1+2k] \\
	&\qquad {} + [1+2i,2j,1+2k] - [1+2i,2j+2,1+2k] \\
	&\qquad\qquad {} - [1+2i,1+2j,2k] + [1+2i,1+2j,2k+2].
\end{align*}
Lest $i=j=k=0$ (in which case the resulting requirement fixes the hitherto unspecified coefficient $a_0 = 12$), it reads
\begin{align}
\begin{split} \label{invariance}
0 &= \beta_0(i,j,k) - \beta_0(i+1,j,k) \\
	&\qquad {} + \beta_0(j,i,k) - \beta_0(j+1,i,k) \\
	&\qquad\qquad {} + \alpha_0(i,j,k) - \alpha_0(i,j,k+1).
\end{split}
\end{align}
If either $i\notin\set{k,k+1}$ or $j\notin\set{k,k+1}$, \autoref{invariance} reads
\[
0 = \alpha_0(i,j,k) - \alpha_0(i,j,k+1)
\]
which holds true (being non-trivial only when $i=j=0$). So, suppose that $i\in\set{k,k+1}$ and $j\in\set{k,k+1}$. If $i=j=k$, \autoref{invariance} reads
\[
0 = -2\beta_0(k+1,k,k) + \alpha_0(k,k,k)
\]
which holds true. If either $i=j-1=k$ or $i-1=j=k$, it reads
\[
0 = -\beta_0(k+1,k+1,k) + \beta_0(k+1,k,k)
\]
which holds true. Finally, if $i=j=k+1$, it reads
\[
0 = 2\beta_0(k+1,k+1,k) - \alpha_0(k+1,k+1,k+1)
\]
which also holds true. Thus, invariance holds.

Now consider the scale $n=-1$.
At this scale, $L$ is specified by the coefficients
\begin{align*}
\alpha_{-1}(i,j,k) &= b_{-1}\bigl([2+4i, 2+4j, 4k]_{-1}\bigr) \\
\beta_{-1}(i,j,k) &= b_{-1}\bigl([4i, 2+4j, 2+4k]_{-1}\bigr)
\end{align*}
with $i,j,k\in \NN$. \autoref{compatibility} enables their computation from those at the scale $n=0$ specified above, and we proceed with it.

Start with $\alpha_{-1}$. Recalling that this is an interaction coefficient between plaquettes parallel to the $(1,2)$-plane,
\[
\alpha_{-1}(i,j,k) = \sum \alpha_0(2i-\delta_1+\delta_3, 2j-\delta_2+\delta_4, 2k)
\]
where the displacements $\delta_1,\dots,\delta_4$ can take values in $\set{0,1}$ and the sum is over all possibilities.
Now, if either $\delta_1-\delta_3$ or $\delta_2-\delta_4$ is not 0, the contribution vanishes. In fact, since $i,j\in\ZZ$, such displacements make it impossible to match any of the patterns $\alpha_0(0,0,k), \alpha_0(k,k,k)$ producing non-zero coefficients.
Thus, we have
\[
\alpha_{-1}(i,j,k) = 4\alpha_0(2i,2j,2k) = 4\alpha_0(i,j,k),\quad i,j,k\in\ZZ.
\]
As for $\beta_{-1}$, one has
\[
\beta_{-1}(i,j,k) = \sum \beta_0(2i-\delta_1, 2j-\delta_2+\delta_3, 2k+\delta_4).
\]
We will see that, again, $\beta_{-1} = 4\beta_0$. We need to check the next four points:
\begin{enumerate}
	\item
$\beta_{-1}(1,0,0) = -8$. Indeed, it equals
\begin{align*}
&\sum \beta_0(2-\delta_1,-\delta_2+\delta_3,\delta_4) \\
	&\qquad= \beta_0(2,1,1) + \beta(2,-1,1) + \beta(1,1,0) + \beta(1,-1,0) + 2\beta(1,0,0) \\
	&\qquad= 2\bigl(\beta(2,1,1) + \beta(1,1,0) + \beta(1,0,0)\bigr).
\end{align*}
	\item
$\beta_{-1}(k+1,k+1,k) = -4$, for all $k\in\NN$. Indeed, it equals
\begin{align*}
	&\sum \beta_0(2k+2-\delta_1,2k+2-\delta_2+\delta_3,2k+\delta_4) \\
	&\qquad= 2\beta_0(2k+2,2k+2,2k+1) + \beta_0(2k+2,2k+1,2k+1) \\
	  &\pushright{{} + \beta_0(2k+1,2k+1,2k).}
\end{align*}
	\item
$\beta_{-1}(k+1,k,k) = -4$, for all $k\geq 1$. Indeed, it equals
\begin{align*}
&\sum \beta_0(2k+2-\delta_1,2k-\delta_2+\delta_3,2k+\delta_4) \\
&\qquad= \beta_0(2k+2,2k+1,2k) + \beta_0(2k+1,2k+1,2k) \\
&\pushright{{} + 2\beta_0(2k+1,2k,2k).}
\end{align*}
	\item
If either $i\neq k+1$ or $j\notin\set{k,k+1}$, then $\beta_{-1}(i,j,k) = 0$. Indeed,
\[
\beta_{-1}(i,j,k) = \sum\beta_0(2i-\delta_1,2j-\delta_2+\delta_3,2k+\delta_4)
\]
and in order for $2i - \delta_1 = 2k + \delta_4 + 1$ one needs $i = k+1$, whereas $2j-\delta_2+\delta_3 \in\set{2k+\delta_4,2k+\delta_4+1}$ requires $j\in\set{k,k+1}$.
\end{enumerate}

Recapitulating, we have proved that
\[
\alpha_{-1}(i,j,k) = 4\alpha_0(i,j,k),\quad \beta_{-1}(i,j,k) = 4\beta_0(i,j,k),
\]
for all $i,j,k\in\NN$. This means that, if we define $L_n$ by
\[
a_n = 4^{-n}a_0,\quad \alpha_n(i,j,k) = 4^{-n}\alpha_0(i,j,k),\quad \beta_n(i,j,k) = 4^{-n}\beta_0(i,j,k),
\]
we get a family of compatible, invariant differential operators, as desired.
\begin{rk}
Other solutions exist.
For instance, just putting $a_0 = 1$ and $b_0(q) = 0$ except for $b_0([1,1,k]) = -1$ if $k\neq 0$ does the job.
\end{rk}

\subsection{$M=\RR^4$ and beyond}

Imposing translation and rotation invariance, we are allowed to specify only the coefficients $b_0(q) = b_{pq}$ with $p=[1,1,0,0]$ and $q$ contained in the principal orthant $\Set{(u_1,\dots,u_4) | u_i\geq 0}$.
By furthermore imposing reflection invariance, we can just focus on plaquettes which are parallel to one of
\[
[1,1,0,0],\ [0,1,1,0],\ [0,0,1,1].
\]
Indeed, under reflection invariance,
\begin{itemize}
\item $b_0([1,0,1,0]) = -b_0([0,1,1,0])$ (the reversed plane contains $p$).
\item $b_0([1,0,0,1]) = b_0([1,0,1,0])$ (the reversed plane does not contain $p$).
\item $b_0([0,1,0,1]) = b_0([0,1,1,0])$ (the reversed plane does not contain $p$).
\end{itemize}
We will write
\begin{align*}
\alpha_0(i,j,k,l) &= b_0([1+2i,1+2j,2k,2l]), \\
\beta_0(i,j,k,l) &= b_0([2i,1+2j,1+2k,2l]), \\
\gamma_0(i,j,k,l) &= b_0([21,2j,1+2k,1+2l]).
\end{align*}
We set $\gamma_0(i,j,k,l)=0$.
As for $\alpha_0$ and $\beta_0$, letting $\alpha_0^{(3)}$ and $\beta_0^{(3)}$ be the coefficients defined in \autoref{R3}, if $k,l\geq 0$ we set
\[
\alpha_0(i,j,k,l) = \alpha_0^{(3)}(i,j,k+l),\quad
\beta_0(i,j,k,l) = \beta_0^{(3)}(i,j,k+l)
\]
and we use reflection invariance to extend them to negative values.
Multiscale consistency holds with the same scaling of the $\RR^3$ case:
\begin{align*}
\alpha_{-1}(i,j,k,l)
	&= \sum \alpha_0(2i-\delta_1+\delta_3, 2j-\delta_2+\delta_4, 2k,2l) \\
	&= \sum \alpha_0^{(3)}(2i-\delta_1+\delta_3, 2j-\delta_2+\delta_4, 2(k+l)) \\
	&= 4\alpha_0^{(3)}(i,j,k+l) = 4\alpha_0(i,j,k,l),
\end{align*}
and similarily for the $\beta$ coefficients. It remains to check gauge invariance.

Take the principal orthant cube $[1+2i,1+2j,1+2k,2l]$. Its boundary is
\begin{align*}
&-[2i,1+2j,1+2k,2l] + [2i+2,1+2j,1+2k,2l] \\
	&\qquad {} + [1+2i,2j,1+2k,2l] - [1+2i,2j+2,1+2k,2l] \\
	&\qquad\qquad {} - [1+2i,1+2j,2k,2l] + [1+2i,1+2j,2k+2,2l].
\end{align*}
Letting $\delta_{ijkl}$ be the Kronecker delta product $\delta_{0i}\delta_{0j}\delta_{0k}\delta_{0l}$, the corresponding invariance requirement reads
\begin{align*}
\delta_{ijkl}a_0 &= \beta_0(i,j,k,l) - \beta_0(i+1,j,k,l) \\
	&\qquad {} + \beta_0(j,i,k,l) - \beta_0(j+1,i,k,l) \\
	&\qquad\qquad {} + \alpha_0(i,j,k,l) - \alpha_0(i,j,k+1,l).
\end{align*}
Since $i,j,k,l\geq 0$, this reduces to the requirement for $[1+2i,1+2j,1+2(k+l)]$ in the case $M=\RR^3$. By reflection invariance, the same restrictions result from $[1+2i,1+2j,2k,1+2l]$.
Consider finally $[2i,1+2j,1+2k,1+2l]$, whose boundary is
\begin{align*}
&-[2i,2j,1+2k,1+2l] + [2i,2j+2,1+2k,1+2l] \\
	&\qquad {} + [2i,1+2j,2k,1+2l] - [2i,1+2j,2k+2,1+2l] \\
	&\qquad\qquad {} - [2i,1+2j,1+2k,2l] + [2i,1+2j,1+2k,2l+2].
\end{align*}
The corresponding restriction reads
\begin{align*}
0 &= \gamma_0(i,j,k,l) - \gamma_0(i,j+1,k,l) \\
	&\qquad {} - \beta_0(i,j,l,k) + \beta_0(i,j,l,k+1) \\
	&\qquad\qquad {} + \beta_0(i,j,k,l) - \beta_0(i,j,k,l+1),
\end{align*}
which holds because $\beta_0(i,j,l,k) = \beta_0(i,j,k,l)$ and
\[
\beta_0(i,j,l,k+1) = \beta_0^{(3)}(i,j,k+l+1) = \beta_0(i,j,k,l+1).
\]
Thus, we have gauge invariance.

At this point, it is clear how to proceed to higher dimensions. We only use $\alpha$ and $\beta$ coefficients, imposing rotation and reflection invariance and defining
\begin{align*}
\alpha_n^{(d)}(i,j,k_1,\dots,k_{d-2}) &= \alpha_n^{(3)}(i,j,k_1+\cdots+k_{d-2}) \\
\beta_n^{(d)}(i,j,k_1,\dots,k_{d-2}) &= \beta_n^{(3)}(i,j,k_1+\cdots+k_{d-2})
\end{align*}
for $k_1,\dots,k_{d-2}\geq 0$.
Multiscale consistency holds just as above, and the new gauge invariance restrictions coming from cubes that do not involve the first two coordinates hold trivially.

\appendix
\section{The Yang-Mills field on $\mathbb S^2$} \label{Yang-Mills}

Fix a cellular decomposition $C$ of $M=\mathbb S^2$, write $C^2M = \set{p_1,\dots,p_n}$ and let $x_i$ stand for the holonomy around $p_i$.
The algebra $A_C$ is
\(
\CC[x]/(x_1+\cdots+x_n),
\)
but we'll need a euclidean coordinate system in this section so we choose to drop $x_n$ in favor of $x_1,\dots,x_{n-1}$.
It is well-known~\cite{levy2003yangmills} that the effective state corresponding to the Yang-Mills field reads, in such coordinates,
\[
\mu_\lambda(f) = Z_\lambda^{-1}\idotsint f\e^{-\sum_{i=1}^n x_i^2/\lambda\abs{p_i}}\D x_1\cdots\D x_{n-1},\quad x_n = -\sum_{i=1}^{n-1}x_i
\]
where
$\abs{p_i}$ is the volume of $p_i$ (we assume that $\sum\abs{p_i}=1$) and $Z_\lambda$ is a normalizing constant.
In this appendix we will provide evidence, by explicitely computing $\tfrac{\D}{\D\lambda} \mu_\lambda$ for small $n$, that
$\mu_\lambda = \mu_0\e^{\lambda L}$ where
\[
L = \sum_{i=1}^{n-1} a_i\partial_i^2 - \sum_{i,j=1}^{n-1} a_ia_j\partial_i\partial_j = \sum_{i=1}^{n} a_i\partial_i^2 - \sum_{i,j=1}^{n} a_ia_j\partial_i\partial_j.
\]
Here, the first expression is in the euclidean coordinate system $(x_1,\dots,x_{n-1})$, whereas the second is in the algebraic coordinate system $\CC[x]/(\sum_{i=1}^n x_i)$.
In order to go from the first to the second, one notes that the algebraic derivation corresponding to the euclidean derivation $\partial_i$ ($i<n$) under the isomorphism
\[ 
x_i\in\CC[x_1,\dots,x_{n-1}] \mapsto x_i\in \CC[x_1,\dots,x_n]/(x_1+\cdots+x_n)
\] 
is $\partial_i - \partial_n$. The latter expression is, thus, the result of making these substitutions in the former (we encourage the reader to do the calculation).

\subsubsection*{Case $n=3$}

Let $x=x_{p_1}$ and $y=x_{p_2}$. Note that $x_{p_3} = -x-y$. Write $a_i=\abs{p_i}$. We want to figure out the first two terms in the power series expansion of
\begin{equation} \label{toy case}
	C_{\lambda}^{-1} \iint f(x)g(y) k_{\lambda a_1}(x) k_{\lambda a_2}(y) k_{\lambda a_3}(x+y)\, \D x\D y
\end{equation}
where $C_t = (4\pi t)^{-1/2}$
and $\set{k_t | t\geq0}$ is the heat kernel semigroup.
Write the integrand as
\[
\frac{C_{\lambda a_3}}{C_{\lambda}} f(x)g(y)\e^{-(x+y)^2/4\lambda a_3} k_{\lambda a_1}(x) k_{\lambda a_2}(y).
\]
In order to develop in power series we absorb the factor $\e^{-(x+y)^2/4\lambda a_3}$ into the heat kernels by a suitable change of variables. We start by writing the integral $\D x$ as follows:
\begin{align*}
&\int f(x) \e^{-(x+y)^2/4\lambda a_3} k_{\lambda a_1}(x)
	= C_{\lambda a_1} \int f(x) \e^{-(x+y)^2/4\lambda a_3} \e^{-x^2/4\lambda a_1} \\
	&\qquad= C_{\lambda a_1} \int f(x) \exp\left( \tfrac{-\left(x+\frac{a_1}{1-a_2}y\right)^2}{ 4\lambda \frac{a_1a_3}{1-a_2} }\right) \exp\left( \tfrac{-y^2}{4\lambda(1-a_2)} \right) \\
	&\qquad=  \frac{C_{\lambda a_1}}{C_{\lambda\frac{a_1a_3}{1-a_2}}} \exp\left( \tfrac{-y^2}{4\lambda(1-a_2)} \right) \int f\left(x-\tfrac{a_1}{1-a_2}y\right) k_{\lambda\frac{a_1a_3}{1-a_2}}(x).
\end{align*}
Therefore, \eqref{toy case} equals
\begin{align*}
&\bigl( 1-a_2 \bigr)^{-1/2} \iint f\left(x-\tfrac{a_1}{1-a_2}y\right)g(y) \exp\left( \tfrac{-y^2}{4\lambda(1-a_2)} \right) k_{\lambda\frac{a_1a_3}{1-a_2}}(x) k_{\lambda a_2}(y) \\
	&\qquad = 
	\iint f\left(x-\tfrac{a_1}{1-a_2}y\right)g(y) k_{\lambda\frac{a_1a_3}{1-a_2}}(x) k_{\lambda a_2(1-a_2)}(y).
\end{align*}
Now we expand in power series using the fact that $k_t = (1+t\Delta +\cdots)\delta$. Start with
\[
	\int f\left(x-\tfrac{a_1}{1-a_2}y\right) k_{\lambda\frac{a_1a_3}{1-a_2}}(x) = \left(1+\lambda \tfrac{a_1a_3}{1-a_2} \Delta\right) \left.f\left(x-\tfrac{a_1}{1-a_2}y\right)\right|_{x=0} + O(\lambda^2).
\]
Thus, for \eqref{toy case} we get, up to order $\lambda$,
\begin{align*}
	&\bigl( 1+\lambda a_2(1-a_2)\Delta \bigr) \left( f\left(-\tfrac{a_1}{1-a_2}y\right) g(y) + \lambda\tfrac{a_1a_3}{1-a_2} \Delta f\left(-\tfrac{a_1}{1-a_2}y\right) g(y) \right)\Bigr|_{y=0} \\
	&\quad = fg\bigr|_0 + \lambda a_2(1-a_2) \left( \bigl( \tfrac{a_1}{1-a_2} \bigr)^2 f''g - \tfrac{2a_1}{1-a_2} f'g' + fg'' \right) \Bigr|_0   + \lambda\tfrac{a_1a_3}{1-a_2} f''g\bigr|_0  \\
	&\quad = fg\bigr|_0 + \lambda \bigl( a_1(1-a_1) f''g + a_2(1-a_2)fg'' - 2a_1a_2f'g'  \bigr) \bigr|_{0.}
\end{align*}
This proves that $\tfrac{\D}{\D\lambda}\mu_\lambda = \mu_\lambda L$ for $\lambda=0$, and a similar calculation works for $\lambda>0$ (one has to start with $k_{\lambda+t} = (1+t\Delta+\cdots)k_\lambda$).

\subsubsection*{Case $n=4$}

Our workhorse in the case $n=3$ was the identity
\[
	\e^{ -(x_1+y)^2/a_n } k_{a_1}(x_1) = \left( \tfrac{a_n}{a_1+a_n}\right)^{1/2} \e^{ -y^2/(a_1+a_n) } k_{\frac{a_1a_n}{a_1+a_n}}\left( x_1 + \tfrac{a_1}{a_1+a_n}y \right)
\]
where, for readability, we have dropped $\lambda$ and some irrelevant factors of 4. This can be  iterated: the integrating weight in the case of 4 plaquettes is
\begin{align*}
	&\frac1{(a_1+\cdots+a_4)^{1/2}} k_{\frac{a_1a_4}{a_1+a_4}}\left( x_1+\tfrac{a_1}{a_1+a_4}(x_2+x_3) \right)  \\*
		&\pushright{ {}\cdot k_{\frac{a_2(a_1+a_4)}{a_1+a_2+a_4}}\left( x_2+\tfrac{a_2}{a_1+a_2+a_4}x_3 \right) k_{\frac{a_3(a_1+a_2+a_4)}{a_1+a_2+a_3+a_4}}(x_3) } \\
	&\quad = k_{\frac{a_1a_4}{a_1+a_4}}\left( x_1+\tfrac{a_1}{a_1+a_4}(x_2+x_3) \right) k_{\frac{a_2(a_1+a_4)}{1-a_3}}\left( x_2+\tfrac{a_2}{1-a_3}x_3 \right) k_{a_3(1-a_3)}(x_3).
\end{align*}
Thus, the expansion we want is given by
\begin{multline*}
\left.\Bigl( 1+ \lambda a_3(1-a_3) \Delta_{x_3} \Bigr)\right|_{x_3=0} \left.\left( 1+ \lambda a_2\tfrac{a_1+a_4}{1-a_3} \Delta_{x_2} \right)\right|_{x_2=-\frac{a_2}{1-a_3}x_3} \\
	\left.\left( 1+ \lambda a_1\tfrac{a_4}{a_1+a_4} \Delta_{x_1} \right)\right|_{x_1=-\frac{a_1}{a_1+a_4}(x_2+x_3)}
\end{multline*}
acting on
$f_1(x_1)f_2(x_2)f_3(x_3)$.
The term of order 0 is $f_1f_2f_3|_{x_i=0}$, while that of order 2 is the sum of the following three terms, evaluated at $x_i=0$:
\[
	\frac{a_1a_4}{a_1+a_4} f_1''f_2f_3\,,
\]
\[
	\frac{a_2(a_1+a_4)}{1-a_3}\left( \frac{a_1^2}{(a_1+a_4)^2} f_1''f_2f_3 - \frac{2a_1}{a_1+a_4} f_1'f_2'f_3 + f_1f_2''f_3 \right),
\]
\begin{multline*}
	a_3(1-a_3)\left( \frac{a_1^2}{(a_1+a_4)^2} \left( 1-\frac{a_2}{1-a_3} \right)^2  f_1''f_2f_3 + \frac{a_2^2}{(1-a_3)^2}f_1 f_2''f_3 \right. \\
	{} + f_1f_2 f_3'' + 2\frac{a_1}{a_1+a_4}\left( 1-\frac{a_2}{1-a_3} \right)\frac{a_2}{1-a_3}  f_1' f_2'f_3  \\
	\left. \vphantom{\left( 1-\frac{a_2}{1-a_3} \right)^2} - 2\frac{a_1}{a_1+a_4}\left( 1-\frac{a_2}{1-a_3} \right) f_1'f_2f_3' - 2\frac{a_2}{1-a_3} f_1f_2' f_3' \right).
\end{multline*}
Let us collect terms.
\begin{itemize}
	\item $f_1f_2 f_3''$ goes with $a_3(1-a_3)$.
	\item $f_1 f_2''f_3$ goes with
\begin{align*}
&\frac{a_2(a_1+a_4)}{1-a_3} + a_3(1-a_3) \frac{a_2^2}{(1-a_3)^2}
	= 	a_2\frac{(1-a_2-a_3)+a_2a_3}{1-a_3} \\
	&\qquad = a_2\frac{(1-a_3) - a_2(1-a_3)}{1-a_3} = a_2(1-a_2).
\end{align*}
	\item $ f_1''f_2f_3$ goes with
\begin{align*}
&\frac{a_1a_4}{a_1+a_4} + \frac{a_1^2a_2}{(a_1+a_4)(1-a_3)} + a_3(1-a_3) \frac{a_1^2}{(1-a_3)^2} \\
	&\qquad= \frac{a_1a_4}{a_1+a_4} + \frac{a_1^2}{1-a_3}\left( \frac{a_2}{a_1+a_4} + a_3 \right) \\
	&\qquad= \frac{a_1a_4}{a_1+a_4} + \frac{a_1^2}{1-a_3}\left( \frac{a_2 + a_3 - a_2a_3 - a_3^2}{a_1+a_4} \right) \\
	&\qquad = \frac{a_1a_4}{a_1+a_4} + \frac{a_1^2}{1-a_3}\frac{(a_2+a_3)(1-a_3)}{a_1+a_4}
	= \frac{a_1}{a_1+a_4}\bigl( a_4 + a_1(a_2+a_3) \bigr) \\
	&\qquad= \frac{a_1}{a_1+a_4}\bigl( a_4+a_1(1-a_1-a_4) \bigr)
	= \frac{a_1}{a_1+a_4}(1-a_1)(a_1+a_4) \\
	&\qquad= a_1(1-a_1).
\end{align*}
	\item $f_1 f_2' f_3'$ goes with $a_3(1-a_3)\frac{-2a_2}{1-a_3} = -2a_2a_3$.
	\item $ f_1'f_2 f_3'$ goes with $-2a_3(1-a_3)\frac{a_1}{1-a_3} = -2a_1a_3$.
	\item $ f_1' f_2'f_3$ goes with
\[
-2\frac{a_1a_2}{1-a_3} + 2a_3(1-a_3)\frac{a_1}{1-a_3} \frac{a_2}{1-a_3}
	= \frac{-2a_1a_2 + 2a_1a_2a_3}{1-a_3} = -2a_1a_2.
\]
\end{itemize}

\begin{rk}
The change of variables being used corresponds to a Cholesky factorizarion. That enables one to make more computations, also for more general cellular complexes, using a computer algebra system.
It also provides enough structure to make it reasonable to attempt a proof by induction, but we will not do so here.
\end{rk}

\bibliographystyle{amsplain}
\bibliography{qft}

\end{document}